\documentclass[12pt,prl,nofootinbib,reprint]{revtex4-1}

\usepackage{amsmath, amssymb, graphicx, color}

\usepackage{hyperref}

\usepackage{times}

\newcommand{\beq}{\begin{equation}}
\newcommand{\eeq}{\end{equation}}
\newcommand{\bes}{\begin{equation*}}
\newcommand{\ees}{\end{equation*}}
\newcommand{\bea}{\begin{align}}
\newcommand{\ena}{\end{align}}

\newcommand{\bra}{\langle}
\newcommand{\ket}{\rangle}

\newcommand{\AdS}{\text{AdS}}
\newcommand{\CFT}{\text{CFT}}



\begin{document}

\title{Holographic entanglement entropy beyond coherent states}
\author{Curtis T. Asplund}
\affiliation{Department of Physics, University of California at Santa Barbara, CA 93106}
\email{casplund@physics.ucsb.edu}

\begin{abstract}
We study entanglement entropy for a class of states in quantum field theory that are entangled superpositions of coherent states with well-separated supports, analogous to Einstein-Podolsky-Rosen or Bell states. We find the contributions beyond the area law. In the case of strongly-coupled conformal field theories, we argue that these states are holographically dual to superpositions of bulk geometries. We conclude that for these states one can use the Ryu-Takayanagi holographic entanglement entropy formula to calculate some terms in the entanglement entropy, but that there can be additional $O(N^{2})$ contributions. We argue that this class of states includes those generated by local quenches and thus that these cannot be described by a classical dual geometry. These considerations may be important for more fine grained treatments of holographic thermalization.
\end{abstract}

\maketitle

\section{Introduction}

The AdS/CFT correspondence \cite{Maldacena:1997re} has been widely used to study
certain strongly-coupled conformal field theories (CFTs) using the dual description 
of certain states in those theories as asymptotically anti-de Sitter (AdS) solutions 
of supergravity. One quantity of interest for CFT states is the entanglement 
entropy between some spatial region $A$ and its complement, and there is a proposal
for how to compute this from the dual geometry \cite{Ryu:2006bv,Hubeny:2007xt} that has by now 
accumulated a large amount of evidence in its favor.
This proposal 
implicitly assumes the existence of a dual geometry.

However, many states of the CFT will have duals that cannot be described 
geometrically. One purpose of this letter is to argue that the state generated by a
local quench of a CFT is such a state.
Local quenches are interesting dynamical processes that have been studied in detail in 
conformal field theory, e.g., \cite{2007JSMTE..10....4C, 2011JSMTE..08..019S, Asplund:2011cq}. 
There have been attempts to find the dual geometry to the state generated by a local 
quench, but none have so far succeeded, which is consistent with our claim 
that no such dual geometry exists.

Local quenches are important because they may offer a way to measure entanglement entropy \cite{2011PhRvL.106o0404C}
and because they offer a localized version of the dynamics responsible for 
thermalization after a global quench \cite{Calabrese:2005in, Calabrese:2006rx, Calabrese:2007rg}.
Such global quenches have the starting point for all extant studies of 
holographic thermalization, see, e.g., \cite{Hubeny:2007xt, Chesler:2008hg, Bhattacharyya:2009uu, AbajoArrastia:2010yt, Albash:2010mv, Das:2010yw, Balasubramanian:2011ur, Chesler:2011ds}. In light of this,
another purpose of this letter is to describe a general class of states
similar to the local quench states 
and to investigate their entanglement entropy, to see how the proposal in \cite{Ryu:2006bv}
might be generalized.


\section{Description of states}

For a given AdS/CFT duality, one expects that there is a large class of states of the CFT that 
are not dual to a classical bulk geometry, in particular states that are (sufficiently)
 non-coherent with respect to their
field expectation values.\footnote{Mixed states of the CFT may have good geometric duals, e.g., black holes,
but these are usually interpreted as coarse-grained dual descriptions.
Here I'm not considering coarse graining beyond tracing over the complement of a spatial region, so
throughout this letter I will only be considering
pure states of the CFT.} This point has been made before, e.g., in \cite{Shepard:2005zc, Balasubramanian:2007zt} 
in the context of the 1/2 BPS sector of the $\mathcal{N} = 4$ SYM theory.
Without attempting to describe exactly which states are dual to a classical bulk
geometry, let us refer to such states as coherent.

We can construct a simple example of a non-coherent state,
	 reminiscent of the Einstein-Podolsky-Rosen 
thought-experiment \cite{PhysRev.47.777}, i.e., 
two well-separated bits of matter in an entangled state.
This is also called a Bell state after the well-known work \cite{Bell:1964kc}. 
To set this up, divide the space of a quantum field theory into two regions. 
For simplicity, consider a one-dimensional space. Denote the regions left and right of the origin as $A$ and $B$, respectively, then let us assume we can write the Hilbert space of the full theory as a tensor product 
$\mathcal{H} = \mathcal{H}_{A}\otimes \mathcal{H}_{B}$.

	 Now let $\phi$ be some quantum field. For simplicity suppose $\phi$ is a real scalar field. Then consider a state $|\phi_{A}^{+}\ket$ of the field corresponding a lump of matter away from the origin. More specifically, consider a coherent 
state of the field operator with a positive Gaussian profile, i.e.,  
$\bra 	\phi_{A}^{+} |\phi(x)|\phi_{A}^{+}\ket = A_{0} e^{-(x+a)^{2}/\sigma^{2}}$. Then we can also consider the state with a negative profile:
$\bra 	\phi_{A}^{-} |\phi(x)|\phi_{A}^{-}\ket = -A_{0} e^{-(x+a)^{2}/\sigma^{2}}$. Similarly for region $B$ we have
$\bra 	\phi_{B}^{\pm} |\phi(x)|\phi_{B}^{\pm}\ket = \pm A_{0} e^{-(x-a)^{2}/\sigma^{2}}$. Well separated means $a \gg \sigma$. 
By taking $A_{0}$ sufficiently large we can ensure that $\bra \phi_{A,B}^{\pm} | \phi_{A,B}^{\mp}\ket \approx 0$
and furthermore that $\bra \phi_{A,B}^{\pm} | O(x) |\phi_{A,B}^{\mp}\ket \approx 0$ where $O(x)$ is 
the field operator, the momentum operator, or the Hamiltonian.
\footnote{We are assuming a conventional field theory that admits coherent states 
and has a Hamiltonian with terms polynomial in the 
field and momentum operators.}

Further consider the states $|\phi_{A}^{\text{s}}\ket = \frac{1}{\sqrt{2}} (|\phi_{A}^{+}\ket + |\phi_{A}^{-}\ket)$ (and analogously defined $|\phi_{B}^{\text{s}}\ket$). Here ``s'' stands for superposed.
We assume that the 
Hamiltonian has a $\phi \rightarrow -\phi$ symmetry so all these states have the same energy.

Now for the total state of the CFT we have options, e.g.,
\begin{align}
	\label{eq:state1}
	\text{Entangled:}\quad &|\psi_{1}\ket = \frac{1}{\sqrt{2}} (|\phi_{A}^{+}\ket|\phi_{B}^{-}\ket + |\phi_{A}^{-}\ket|\phi_{B}^{+}\ket) \\
\label{eq:state2}
	\text{Unentangled:} \quad &|\psi_{2}\ket = |\phi_{A}^{\text{s}}\ket|\phi_{B}^{\text{s}}\ket.
\end{align}
Both states have the same energy and, in fact, 
the same stress-tensor expectation value $\bra T_{\mu\nu}(x)\ket$ and 
the same expectation values for the field, $\bra \phi(x) \ket = 0$, for all $x$,
for large $A_{0}$.
However, we would expect that state 1 has roughly $\log 2$ more entanglement entropy
than state 2. 
Below we will verify this expectation in a simple model where we can explicitly
compute the entanglement entropy,
including area law contributions.

In a holographic context, we may consider a CFT that has a holographic dual, 
for example the canonical $\mathcal{N} = 4$, four-dimensional $\text{SU}(N)$ gauge theory.
The large $N$, large coupling limit of the CFT is conjectured to be dual to type IIB supergravity 
on an asymptotically $\AdS_{5}\times S^{5}$ spacetime 
\cite{Maldacena:1997re}. There is an analogous conjectured duality between
certain two-dimesional CFTs and asymptotically $\AdS_{3}$ spactimes.

In the case of coherent states of the CFT that can be matched with a particular geometry,
i.e., a solution to the appropriate supergravity theory, there is a proposal for 
how to calculate the entanglement entropy of the CFT state. 
For a spatial region $A$, the entanglement entropy $S_{A}$ of the reduced density matrix
$\rho_{A}$ (traced over the complement of $A$) is given by
\beq
\label{eq:RT}
	S_{A} = \frac{\text{Area}(\gamma_{A})}{4 G_{\text{N}}},
\eeq
where $\gamma_{A}$ is the extremal surface of minimal area 
in the bulk whose boundary coincides with the boundary of $A$
\cite{Ryu:2006bv,Hubeny:2007xt}. 
In the duality $G_{N}^{-1}$ goes like $N^{2}$, so in the large $N$ 
limit the above formula will only capture $O(N^{2})$ contributions.

In this context there are states analogous to those in \eqref{eq:state1} 
and \eqref{eq:state2} that we can obtain by superposing coherent traveling excitations
in the CFT. States describing a single traveling excitation, a ``pulse,'' and their 
holographic duals were recently 
described in  \cite{Roberts:2012aq}, in $\AdS_{3}/\CFT_{2}$.
The states in \cite{Roberts:2012aq} were mixed states of the CFT, but there 
are corresponding pure states with the same essential properties, i.e.,
supported in a finite region of the null coordinates and with 
$O(N^{2})$ of the fields excited.
We can also consider two left-moving excitations that
are separated into the left and right regions of the CFT for some period of time, see Fig. 1.


\begin{figure}[h]
\begin{center}
\includegraphics[width=8cm]{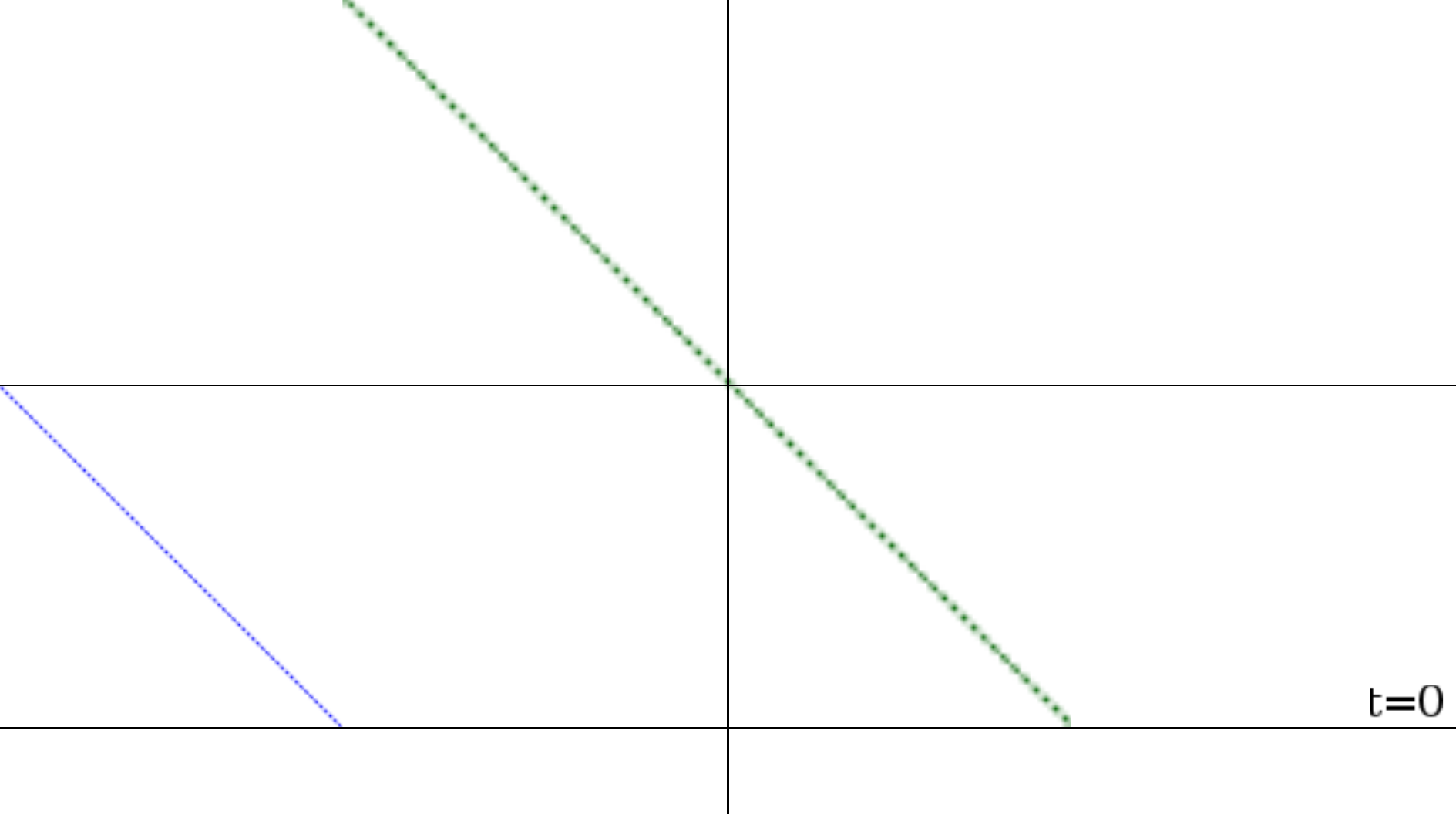}
\end{center}
\caption{Two left moving excitations in the CFT, which may be entangled or not.}
 \label{fig:pulses}
\end{figure}


We can construct pure states corresponding to the pulses either being 
entangled or not, in analogy with \eqref{eq:state1} and \eqref{eq:state2} above.
We expect the maximally entangled state, call it $|\psi_{1}\ket$, will have 
$O(N^{2})$ more entanglement entropy than the unentangled state $|\psi_{2}\ket$.
We will corroborate this expectation with an explicit calculation in a toy model
below. Let us also note that the maximally entangled state is similar to that
investigated in \cite{Asplund:2011cq}, which was produced after a local quench. The entanglement 
entropy was found there to be proportional to the central charge, which would be 
$O(N^{2})$ in a holographic theory.

We can argue, by contradiction, that theses states $|\psi_{1,2}\ket$ do not have a dual bulk geometry.  
The first premise is that the entangled state
will have $O(N^{2})$ more entanglement entropy than the unentangled state.
The second is that, assuming these 
two states had a dual bulk geometry, they would be same. 
The reason for this is that these two states are identical as far as $\bra T_{\mu\nu}(x) \ket$ and the expectation 
values of the fields, to $O(N^{2})$,
so this is what we expect from the usual AdS/CFT dictionary \cite{Witten:1998qj, Balasubramanian:1998sn}. 
Then, using \eqref{eq:RT} we would find that the $O(N^{2})$ entanglement 
entropy would be the same, a contradiction.

The point of the above argument is just to 
emphasize that simple superpositions of coherent states are in the 
class of states that don't have a dual bulk geometry.
Such simple superpositions are interesting for at least two reasons. First, 
while \eqref{eq:RT} is not directly applicable, it may still provide some 
terms in the full expression for the entanglement entropy of certain superpositions, 
as we discuss further in the Conclusion. Second, states of this kind are 
produced by quenches, as I will argue later, and may be important in 
general dynamical situations.

\section{Model for calculating entanglement entropy}
\label{sec:model}

We can investigate some of these issues explicitly by considering a toy model of 
the ground state of a free quantum field theory.
The point of this toy model is only to capture the leading contributions to 
the entanglement entropy.
From early investigations of the area law \cite{Srednicki:1993im, Bombelli:1986rw}, the entanglement entropy of a 
region of space in the ground state of a field has been understood to come from
correlations across the boundary of the region. This explains why the entropy 
goes as the area and why it diverges in the absence of a UV cutoff---arbitrarily 
small distances near the boundary can contribute.
\footnote{This explanation is only compelling in the case of fields whose correlation
length is small compared to the dimension of the region, i.e., sufficiently massive 
fields. However, the area law is still gives the leading order behavior in the massless case 
\cite{Srednicki:1993im} and 
so our model should still be relevant for that case. }

 We can model this situation by considering a discrete space. For now just take the
 radial direction to be discrete, indexed by $i$ so $r = ia$ where $a$ is the 
lattice spacing.
Decompose the Hilbert space of the theory into a tensor product, one factor 
for each $i$.
Then the entanglement between different points in the space can be roughly modeled
with a state of the form
\beq
	|0\ket \approx  \left(\sum_{i=1}^{i_{\text{max}}-1}e^{a_{i}^{\dagger}a_{i+1}^{\dagger}} \right)\prod_{i=1}^{i_{\text{max}}}|0\ket_{i},
\eeq
where $i_{\text{max}}a$ is some IR cutoff, not important for our purposes here.
This exhibits entanglement between nearest neighbor radii. Tracing over
the region insides a sphere of radius $R$ corresponds to tracing over factors up to 
and including
$i_{R} = R/a$ and the only contribution to the entanglement entropy would be from the 
term containing $\exp\left(a_{i_{R}}^{\dagger}a_{i_{R}+1}^{\dagger}\right)|0\ket_{i_{R}}|0\ket_{i_{R}+1}$.
The trace over the $i_{R}$ factor would result in a reduced density matrix for the $i_{R} + 1$ that was
mixed and in fact thermal, i.e., of the form $\sum_{\alpha} e^{-\beta E_{\alpha}} |E_{\alpha}\ket_{i_{R}+1}\bra E_{\alpha}|_{i_{R}+1}$. 
The corresponding entanglement entropy would thus be the von Neumann entropy of a 
thermal state on a sphere, which we expect to be extensive with respect to 
the area of the sphere, and so we get an area law: $S_{|0\ket} \propto R^{d-1}$,
if the dimension of space is $d$. 
The above model is clearly quite crude and only captures the leading, area-law behavior
of the entanglement entropy. However this is the leading order behavior 
even in the vacuum of the strongly-coupled field theory \cite{Nishioka:2009un}.

Now let us consider an excited state which contains entanglement besides the nearest 
neighbor type considered above. We can model this with states 
\bea
\label{eq:model-states}
	|\psi_{\pm} \ket &= C_{1} e^{\pm\lambda a_{1}^{\dagger}\mp \lambda a_{j}^{\dagger}} |0\ket \\
	 &= \left(\sum_{i=1}^{i_{\text{max}}-1}e^{a_{i}^{\dagger}a_{i+1}^{\dagger}} \right)| {\pm}\lambda\ket_{1} |0\ket_{2}\cdots |{\mp}\lambda\ket_{j}\cdots, 
\end{align}
where $\lambda > 0$ and $j\gg 1$. The $|\pm \lambda\ket$ are coherent states.
These are crude models for the $|\phi^{\pm}\ket$ considered above. 
Each of the states $|\psi_{\pm} \ket$ are unchanged from the vacuum in terms 
of their entanglement entropy: they will each obey an area law. 
We now 
want to consider a superposition of these coherent states 
\beq
\label{eq:ent-state}
|\psi\ket = C_{1} |\psi_{+}\ket + C_{2}|\psi_{-}\ket,
\eeq where for simplicity
assume that the coefficients $C_{i}$ are real, and a trace over 
radii smaller than $ja$. The resulting reduced density matrix is
\bea
\nonumber
	\rho_{\text{red}}& = C_{1}^{2} \left(\sum_{\alpha} e^{-\beta E_{\alpha}} |E_{\alpha}\ket_{i_{R}+1}\bra E_{\alpha}|_{i_{R}+1}\right)
	\cdots |{-}\lambda\ket_{j}\bra {-}\lambda|_{j}\cdots \\
	&+ C_{2}^{2} \left(\sum_{\alpha} e^{-\beta E_{\alpha}} |E_{\alpha}\ket_{i_{R}+1}\bra E_{\alpha}|_{i_{R}+1}\right)
	\cdots |{+}\lambda\ket_{j}\bra {+}\lambda|_{j}\cdots.
\end{align}
Luckily this form is diagonal (choosing a coherent state basis for the $j$th factor), and we can compute
the entanglement entropy. We get 
\beq
	S_{|\psi\ket} = S_{|0\ket} - \gamma \log \gamma - (1-\gamma)\log (1-\gamma),
\eeq
where $\gamma = C_{1}^{2}$. For small $\gamma$ this is
\beq
		S_{|\psi\ket} \approx S_{|0\ket} - \gamma \log \gamma.
\eeq
Of course, if we had chosen an entangling region such that $R > ja$, then the entanglement 
entropy would return to its vacuum value.

We can generalize this to some number $M$  fields $\phi_{k}$. Denote the coherent state of field
$k$ at radial position $i$ by $|\lambda\ket_{k,i}$. Now we imagine that all of these fields
are excited near the origin and at radial position $ja$. One example of such a state is
\bea
|\psi_{(\pm\cdots \pm)}\ket &= |\pm\lambda\ket_{1,1}|0\ket_{1,2}\cdots |\mp\lambda\ket_{1,j}\cdots \\
\nonumber	&\otimes |\pm\lambda\ket_{2,1}|0\ket_{2,2}\cdots |\mp\lambda\ket_{2,j}\cdots \\
\nonumber	&\cdots \\
\nonumber	&\otimes |\pm\lambda\ket_{M,1}|0\ket_{M,2}\cdots |\mp\lambda\ket_{M,j}\cdots,
\end{align}
where we have suppressed the nearest-radial-neighbor entanglement operators for brevity.
The index $\beta = (\pm\cdots \pm)$ runs over the $2^{M}$ possible states. 
The maximal 
entropy such state is
\beq
\label{eq:maxE}
|\psi\ket = \frac{1}{\sqrt{2^{M}}} \sum_{\beta} |\psi_{(\pm\cdots \pm)}\ket.
\eeq
If we again trace over radii smaller than $ja$ we find an entanglement entropy
\beq
\label{eq:Smodel}
	S_{|\psi\ket} = S_{|0\ket} + M \log 2.
\eeq
It's also easy to construct an unentangled superposition 
by analogy with \eqref{eq:state2}, by using states
$|\psi^{\text{s}}\ket_{1,k} = 2^{-1/2}(|{+}\lambda\ket_{1,k}
+ |{-}\lambda\ket_{1,k})$. This will share with the state in \eqref{eq:maxE}
 the property that $\bra \phi_{k}(x) \ket =0$ 
for each field, but will have entanglement entropy equal to that of the vacuum.

%

\section{Discussion}


%

We now consider the implications for the holographic entanglement entropy proposal.
Can we compute the entanglement entropy for such non-coherent states of the boundary CFT, 
of the kind discussed above, from bulk information?
The simple model considered above suggests that the leading order 
entanglement entropy for the simplest such state in the CFT is given by 
\beq
\label{eq:corr}
	S_{A} \approx \frac{\text{Area}(\gamma_{A})}{4 G_{\text{N}}} + N^{2} C \log 2,
\eeq
where $C$ is a constant of order $1$, the fraction of the $O(N^{2})$ fields
that are entangled.
This does not tell us how to compute the entanglement entropy from the bulk, 
rather it is a constraint 
on any proposal for computing $S_{A}$ from bulk data for a states of this kind.
However, from the bulk point of view, 
the state might be interpreted as a superposition of geometries. 

We can consider a slightly more general situation. 
First, if the state of your boundary
CFT is in a (sufficiently) coherent state, then you can use the original \eqref{eq:RT}.
If not, write your state in a basis of coherent states. 
For simplicity let us consider the simplest case
of a state that describes $O(N^{2})$ fields 
supported in two regions that are well separated and in the maximally 
entangled state.
If your entangling surface separates those regions, then the model 
suggests the conjecture that
 the entanglement entropy is given by \eqref{eq:corr}, to leading order in $N$,
where $\text{Area}(\gamma_{A})$ is computed in pure AdS.

This all refers to the special case of superpositions of localized coherent states.
We expect the bulk geometries dual to these various coherent states to be the same throughout 
most of the spacetime and this is why \eqref{eq:RT} might still be used to compute
a term in the full expression for the entanglement entropy.
A general, non-coherent state may be dual to a superposition of wildly different bulk geometries
and wildly different field configurations
and it would be interesting to find generalizations of \eqref{eq:corr} for such states, if such exist.

The additional term appearing in \eqref{eq:Smodel} can be identified with the
entanglement entropy produced following a local quench in a general CFT. 
Though it is difficult to directly access the state of the CFT after the local quench
(although see \cite{Avery:2010er, Avery:2010hs}), studies of entanglement
entropy after a local quench \cite{2007JSMTE..10....4C, 2011JSMTE..08..019S, Asplund:2011cq} suggest
that the state consists of entangled coherent particles emitted from the quench point(s) and 
then propagating at the speed of light. The quench thus creates a state much like that in \eqref{eq:maxE}.
Such a quench, which corresponds to a topology change, is generally expected to excite every field in the theory 
\cite{Anderson:1986ww}.
This is corroborated by the fact that the entanglement entropy produced by a local quench
in 2D CFTs is proportional to the central charge \cite{2007JSMTE..10....4C, 2011JSMTE..08..019S, Asplund:2011cq}.
In the 
holographic context this would mean $O(N^{2})$ contributions to the entanglement entropy of
the form \eqref{eq:corr}.
 Thus, we do not expect 
there to be any classical bulk geometry dual to a local quench of the boundary CFT.

These considerations may be important for holographic thermalization, see, e.g., 
\cite{Hubeny:2007xt, Chesler:2008hg, AbajoArrastia:2010yt, Bhattacharyya:2009uu, Albash:2010mv, Das:2010yw, Balasubramanian:2011ur, Chesler:2011ds}.
Like local quenches, global quenches 
in $1+1$ dimensions
are known to produce pairs of entangled particles that 
can become separated by arbitrarily large distances \cite{Calabrese:2005in, Calabrese:2006rx, Calabrese:2007rg}.
This picture seems to hold at strong coupling and in higher dimensions as well, at least
for the entanglement entropy of a single region \cite{Balasubramanian:2011ur}.
The spatially homogenous nature of the global quench means that one will not get non-coherent states of the kind
considered here, and the classical dual geometries commonly used (such as the Vaidya metric)
seem to offer accurate coarse grained descriptions.
However, more general thermalization scenarios may lead to very non-coherent states,
and so modeling such a process by a single 
dual geometry may not be correct, even to leading order in $N$,
 since the resulting calculations of entanglement entropy may be
missing terms such as those appearing in \eqref{eq:corr}. 
We hope to study this in more detail in future work.

We note that there has already been a holographic study of an entangled pair of spatially
separated subsystems, in \cite{Fujita:2011fp}, Sec. 7.1.
In that system the entanglement in the vacuum state could be understood as an Unruh effect due to the 
fact that the CFT was undergoing uniform acceleration.

We acknowledge that the model we presented is crude and there
are many avenues for further investigation. It would 
be nice to improve the model so as to capture the subleading corrections (in $R/a$) to the area law for states of 
the kind we are considering, to see how they compare. We should also investigate the mutual information
and the tripartite information, since these show departures from the entangled particle picture of
thermalization after a global quench in the strongly coupled case \cite{Balasubramanian:2011at}.
We have been vague about how sufficiently coherent a state needs to be in order to apply the 
holographic entanglement entropy proposal. It would interesting to quantify this and 
study departure from coherence in a dynamical situation and the resulting behavior of the 
entanglement entropy and other quantities.

\begin{acknowledgments} 
  I thank S. Avery, V. Balasubramanian, D. Berenstein, B. Craps,  T. Hertog, D. Marolf, E. Perkins,  M. Roberts, 
  V. Rosenhaus, M. Srednicki and M. Van Raamsdonk
 for all their helpful conversations about this work, and W. Kelly, M. Headrick, G. Horowitz and 
 T. Takayanagi additionally for their 
 comments on an early draft. The author was supported in part by the 
 Department of Energy under Contract DE-FG02-91ER40618, and 
 part of this work was carried out while participating in the KITP program ``Bits, branes, and black holes'' and the associated conference ``Black holes and information'' and was supported in part by the National Science Foundation under Grant No. NSF PHY11-25915. 
 
\end{acknowledgments}

%

%

\end{document}